\documentclass[aps,prl,twocolumn,superscriptaddress]{revtex4}
\usepackage{amsmath}
\usepackage{graphicx}
\usepackage{float}
\usepackage{color}
\usepackage{multirow}
\usepackage{longtable}

\begin{document}


\title{Polarization components in $\pi^{0}$ photoproduction at photon energies up to 5.6 GeV}


\author{W. Luo}
\affiliation{Lanzhou University, Lanzhou 730000, Gansu, People's Republic of China}

\author{E. J. Brash}
\affiliation{Christopher Newport University, Newport News, Virginia 23606, USA}
\affiliation{Thomas Jefferson National Accelerator Facility, Newport News, Virginia 23606, USA}
\author{R. Gilman}
\affiliation{Thomas Jefferson National Accelerator Facility, Newport News, Virginia 23606, USA}
\affiliation{Rutgers, The State University of New Jersey,  Piscataway, New Jersey 08855, USA}
\author{M. K. Jones}
\affiliation{Thomas Jefferson National Accelerator Facility, Newport News, Virginia 23606, USA}
\author{M. Meziane}
\author{L. Pentchev}
\author{C. F. Perdrisat}
\affiliation{The College of William and Mary, Williamsburg, Virginia 23187, USA}
\author{A. J. R. Puckett}
\affiliation{Massachusetts Institute of Technology, Cambridge, Massachusetts 02139, USA}
\affiliation{Los Alamos National Laboratory, Los Alamos, New Mexico 87545, USA}
\author{V. Punjabi}
\author{F. R. Wesselmann}
\affiliation{Norfolk State University, Norfolk, Virginia 23504, USA}
% ...

\date{\today}

\begin{abstract}

We present new data for the polarization observables of the final state proton in the 
$^{1}H(\vec{\gamma},\vec{p})\pi^{0}$ reaction. 
These data can be used to test predictions based on hadron helicity conservation (HHC)
and perturbative QCD (pQCD). 
These data have both small statistical and systematic uncertainties,
and were obtained with beam energies between 1.8 and 5.6 GeV and for $\pi^{0}$ scattering 
angles larger than 75$^{\circ}$ in center-of-mass (c.m.) frame. The data extend the polarization measurements data base  
for neutral pion photoproduction up to $E_{\gamma}=5.6 \mbox{ GeV}$. 
The results show non-zero induced polarization above the resonance region. 
The polarization transfer components vary rapidly with the photon energy and $\pi^{0}$ 
scattering angle in c.m. frame.
This indicates that HHC does not hold and that the pQCD 
limit is still not reached in the energy regime of this experiment.


\end{abstract}

\pacs{}

\maketitle

\paragraph{}


One of the major goals of nuclear physics is to understand the mechanism of exclusive reactions, like meson photoproduction.
Nuclear reactions are described by meson-exchange models at low energy,
and pQCD is expected to apply at very high energy.  
The reaction dynamics of the transition remain unclear in the intermediate energy regime.
The constituent counting rule (CCR) \cite{Brodsky1973} and HHC \cite{Brodsky1981} can be
considered as indications of the applicability of pQCD. 
A scaling behavior for a variety of differential cross sections has been observed for many 
exclusive reactions \cite{Anderson1976,Bochna1998,Belz1995,Freedman1993,Schulte2001,PhysRevLett.91.022003, PhysRevLett.103.012301} 
as predicted by the CCR. But the onset of scaling sometimes starts at the surprisingly low energy of 1 GeV where pQCD should not work.
The limited experimental data do not support the validity of HHC in the few GeV regime. 
Another unsolved problem is that although quark models explain well the baryon excitation states below 2 GeV, these theories also
predict a large density of resonance states at higher energy which have 
not been observed yet \cite{RevModPhys.82.1095}. 
Measurements of both cross section and polarization observables help the understanding
of the dynamics of exclusive reactions.

Prominent structures in the cross section data indicate that $\pi^{0}$ 
photoproduction is dominated by the excitation of baryon resonances at low photon energies, 
E$_{\gamma}<1.8 $ GeV. 
Above the known resonance region, the cross section becomes 
structureless and approximately follows CCR.
Two observables, the induced recoil proton polarization $P$ and the linearly polarized photon asymmetry $\Sigma$,
which are well characterized below 1.5 GeV, provide further evidence of the dominance of resonance excitation
in the E$_{\gamma} < 1.8$ GeV region.
A Jefferson Lab Hall A experiment \cite{gilman02} has obtained data for the three recoil proton polarization components 
and confirmed the importance of polarization observables as a powerful tool in the search for resonance states.
The contribution of these polarization results in constraining multipole analyses was investigated in Ref. \cite{PhysRevC.67.048201},
the conclusion was that more data were needed to constrain the multipoles above 1 GeV.  
The structureless cross section data do not rule out the possibility of overlapping high-mass resonance 
states with large width. 
High precision measurement of polarization observables may give hints to the existence of missing baryon resonance states.

As a consequence of pQCD, and 
with the assumption that orbital angular momentum can be neglected, 
HHC predicts that the 
polarization components of the proton above the baryon resonance region should have a smooth dependence on
 $E_{\gamma}$ and approach limits established by HHC in the absence of baryon resonances 
in the $^{1}H(\vec{\gamma},\vec{p})\pi^0$ reaction. The results from \cite{gilman02} have demonstrated that HHC is not 
valid up to 3 GeV. 
Huang \textit{et al.} \cite{Huang:2003jd} calculated the polarization observables in pion photoproduction 
by assuming the handbag mechanism. However, the theoretical calculation could not be compared to the data because the
photon energy of \cite{gilman02} is not high enough for
the handbag mechanism to be applicable. 
In the past several years it has become increasingly apparent that orbital
angular momentum (OAM) cannot generally be neglected in high energy reactions \cite{PhysRevLett.90.241601, PhysRevC.71.032201}.
This has led to an extension of the CCR to include OAM effects, but to date
there are no predictions for the effects of OAM on polarization observables.
In the absence of resonances, any energy dependence is likely small, but
strong angle variations might persist.
The present work measured the three polarization observables in high precision up to 5.6 GeV.
\begin{figure} 
  \begin{center}
    \includegraphics[width=.5\textwidth, height=.35\textwidth, angle=0,trim = 5mm 2mm 20mm 5mm, clip]{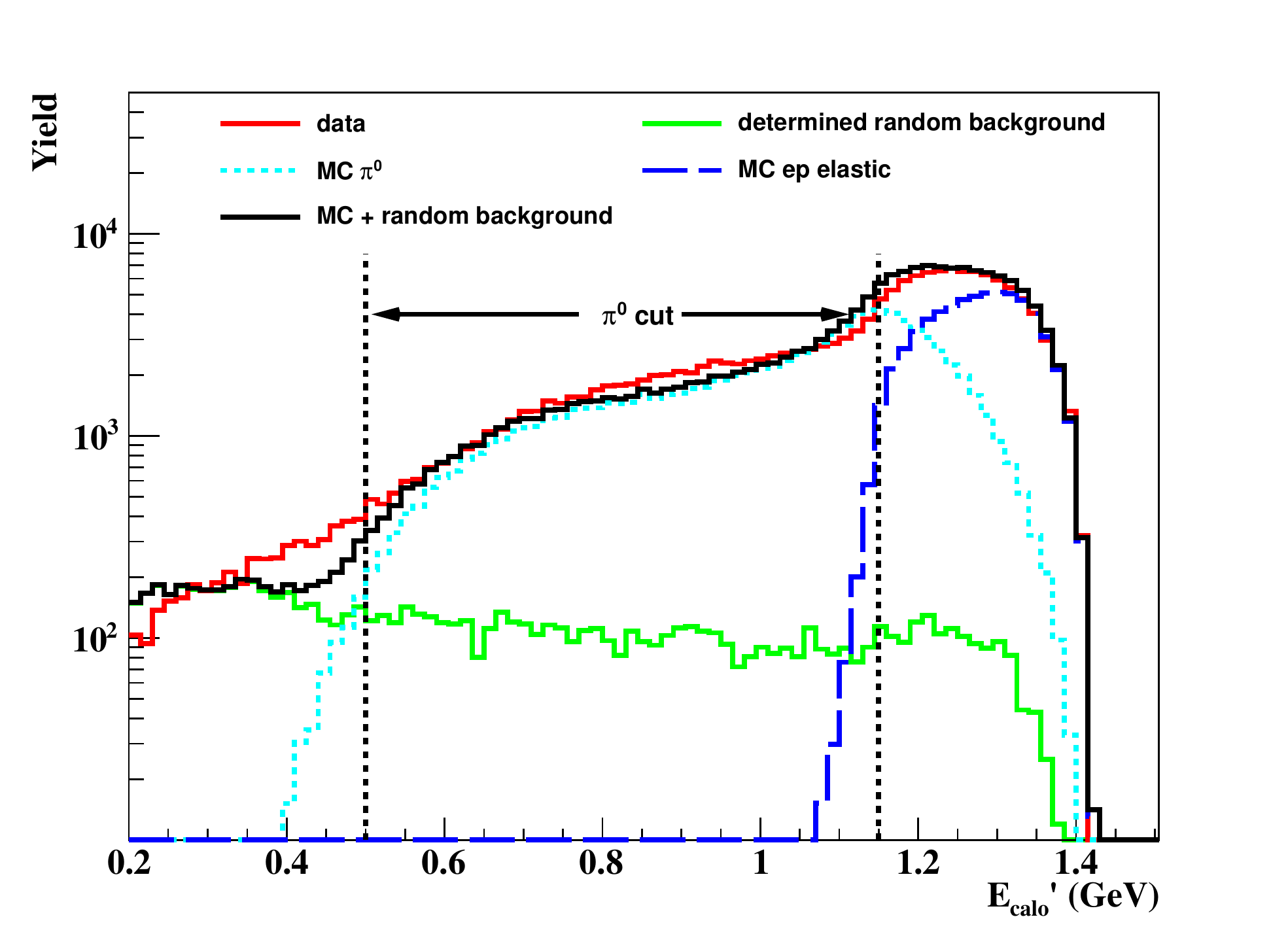}
  \end{center}
  \caption{\label{pi0id} 
The $\pi^{0}$ event selection at an incident photon energy $E_{\gamma} = 3.951 $ GeV.  
The distributions of the predicted energy deposition in the BigCal are plotted for the data (red solid line), 
the random background (green solid line) determined from time of flight spectra,
the Monte Carlo simulation of $\pi^{0}$ events (light blue dotted line),
and the MC simulation of \textit{ep} elastic events (blue dashed line).
The black solid line is the sum of the MC simulation of $\pi^{0}$s, \textit{ep} elastic events and  the measured random background.
The simulated curves have been scaled to match the data. 
The two vertical lines are described in the text.}
\end{figure}

Two experiments were carried out by the GEp-III and GEp-2$\gamma$ collaborations in Hall
 C at Jefferson Lab.
GEp-III measured the elastic proton form factor ratio to high four-momentum transfer, $Q^{2}$, using 
the recoil polarization method in the \textit{ep} elastic reaction \cite{Puckett10}.
GEp-2$\gamma$ measured the kinematic dependence of the ratio at fixed $Q^{2}$ \cite{Meziane11}.
 Due to its relatively 
larger cross section at high $Q^{2}$ and kinematical similarity in phase space to the \textit{ep} elastic
 reaction, neutral pion production was the major contribution to the background of 
these experiments.
The other reactions are suppressed by the \textit{ep} elastic kinematic settings. 
These pions come from real photoproduction as well as electroproduction.
The angular and energy selectivity of these experiments restricted the contribution of 
electroproduction to very low values of $Q^{2}$, \textit{i.e.}, quasi-real photons,
resulting in final states indistinguishable from photoproduction induced by real Bremsstrahlung photons.
Therefore, the polarization observables of the protons in these two reactions are similar 
as proven by a previous experiment \cite{gilman02}. 
In this paper, these two reaction channels are not distinguished and are collectively called neutral pion photoproduction.

\begin{table*} 
\caption{The proton polarization components for the process $^{1}H(\vec{\gamma},\vec{p})\pi^{0}$. 
The $E_{\gamma}$ is the incident photon energy calculated by the proton angle and momentum,
 $\theta^{c.m.}_{\pi^{0}} $ is the angle of $\pi^{0}$ in c.m. frame for each bin of $E_{\gamma}$,  $\chi$ is the proton spin precession angle inside the HMS.
\label{result_table} }
 
\begin{ruledtabular}
\begin{tabular}{  c c c  c c c } 
$\overline{E_{\gamma}} $ (GeV)&  $\overline{\theta^{c.m.}_{\pi^{0}}} $ (deg) &  $\overline{\chi}$ (deg) &$C_{x}^{lab} $ $\pm$ stat. $\pm$ syst. & $C_{z}^{lab}$ $\pm$ stat. $\pm$ syst.	       &$P$  $\pm$ stat. $\pm$ syst.                      \\ \hline

1.845 $\pm$ 0.038 &    143.3  $\pm$  2.5  & 108.9                    &  0.331  $\pm$  0.003 $\pm$  0.006   &   0.073  $\pm$  0.006 $\pm$ 0.005   & -0.503   $\pm$  0.014 $\pm$ 0.012 \\ \hline

2.704 $\pm$ 0.050 &     97.1  $\pm$  2.3  & \multirow{2}{*}{108.9}   &  0.508  $\pm$  0.007 $\pm$  0.005   &   0.255  $\pm$  0.013 $\pm$ 0.004   &  0.138   $\pm$  0.030 $\pm$ 0.009 \\
2.776 $\pm$ 0.025 &     96.1  $\pm$  2.3  &                          &  0.465  $\pm$  0.009 $\pm$  0.005   &   0.263  $\pm$  0.017 $\pm$ 0.003   &  0.023   $\pm$  0.036 $\pm$ 0.009 \\ \hline

3.304 $\pm$ 0.050 &     82.5  $\pm$  2.3  & \multirow{4}{*}{108.9}   &  0.082  $\pm$  0.014 $\pm$  0.009   &   0.358  $\pm$  0.024 $\pm$ 0.009   &  0.215   $\pm$  0.053 $\pm$ 0.014 \\
3.402 $\pm$ 0.050 &     81.6  $\pm$  2.5  &                          &  0.074  $\pm$  0.008 $\pm$  0.009   &   0.362  $\pm$  0.014 $\pm$ 0.009   &  0.210   $\pm$  0.030 $\pm$ 0.012 \\
3.498 $\pm$ 0.050 &     79.7  $\pm$  2.3  &                          &  0.080  $\pm$  0.009 $\pm$  0.008   &   0.343  $\pm$  0.016 $\pm$ 0.008   &  0.151   $\pm$  0.034 $\pm$ 0.010 \\
3.569 $\pm$ 0.030 &	79.4  $\pm$  2.5  &                          &  0.094  $\pm$  0.018 $\pm$  0.008   &   0.293  $\pm$  0.031 $\pm$ 0.010	&  0.237   $\pm$  0.066 $\pm$ 0.012  \\ \hline

3.858 $\pm$ 0.050 &    124.7  $\pm$  4.2  & \multirow{2}{*}{176.0}   &  0.061  $\pm$  0.024 $\pm$  0.007   &   0.742  $\pm$  0.077 $\pm$ 0.020	& -0.176   $\pm$  0.020 $\pm$ 0.011 \\
3.951 $\pm$ 0.050 &    123.3  $\pm$  4.6  &                          &  0.064  $\pm$  0.018 $\pm$  0.003   &   0.699  $\pm$  0.057 $\pm$ 0.018	& -0.174   $\pm$  0.015 $\pm$ 0.009 \\ \hline

5.550 $\pm$ 0.050 &    112.6  $\pm$  4.0  & \multirow{2}{*}{219.5}   &  0.098  $\pm$  0.041 $\pm$  0.007   &  -0.078  $\pm$  0.080 $\pm$ 0.009	&  0.387   $\pm$  0.053 $\pm$ 0.034 \\
5.631 $\pm$ 0.030 &    112.2  $\pm$  5.3  &                          &  0.025  $\pm$  0.054 $\pm$  0.002   &  -0.162  $\pm$  0.104 $\pm$ 0.009	&  0.347   $\pm$  0.070 $\pm$ 0.033 \\ \hline

5.552 $\pm$ 0.050 &    138.1  $\pm$  4.0  & \multirow{2}{*}{261.6}   &  0.198  $\pm$  0.015 $\pm$  0.021   &   0.732  $\pm$  0.016 $\pm$ 0.026	&	 \\ 
5.643 $\pm$ 0.040 &    137.3  $\pm$  5.3  &                          &  0.189  $\pm$  0.016 $\pm$  0.009   &   0.772  $\pm$  0.017 $\pm$ 0.019	&	 \\

\end{tabular}

\end{ruledtabular}
\end{table*}

A high luminosity longitudinally polarized electron beam (79-86\% polarization) was 
scattered from a 20 cm liquid hydrogen target.
In the six kinematic settings of the experiments, the incident electron energy was 1.87, 2.84, 3.63, 4.05 and 5.71 GeV (two settings with $E_{e} =$ 5.71 GeV). 
The beam helicity was flipped at 30 Hz. The beam polarization
 was monitored by the Hall C M\o ller polarimeter \cite{MollerNIM} with an accuracy of 1.0\%.
Near the endpoint, the circular polarization of the Bremsstrahlung photons is nearly equal to 
the longitudinal polarization of the incident electron, 
while the linear polarization component vanishes \cite{Olsen1958}.

The scattered protons were detected in the Hall C High Momentum Spectrometer (HMS) \cite{Blok2008}. 
The proton trajectories were measured by drift chambers in the HMS focal plane. 
The polarization of the proton was measured by the Focal Plane Polarimeter (FPP) in the HMS detector hut 
 downstream from the HMS drift chambers. The FPP, consisting of two 55 cm CH$_{2}$ analyzer
 blocks, each followed by a pair of drift chambers, measured the asymmetry of the charged 
particles in $\vec{p}$+CH$_{2}$$\rightarrow$ charged particle + X to extract the proton polarization.

An electromagnetic calorimeter (BigCal), with a front area of
 1.2 $\times$ 2.2 m$^{2}$, and consisting of 1744 4$\times$4 cm$^2$ lead-glass blocks,
 was placed at the six positions matching the acceptance of the HMS for the elastic $ep$ reaction.
 BigCal provides no discrimination between electrons and photons and gives the 
impact position with similar resolution for both. The BigCal energy resolution changed from 10\%/$\sqrt{E}$ to 23\%/$\sqrt{E}$ 
during the experiment due to radiation damage.
By contrast, the coordinate resolution of about 8 mm is not measurably affected by radiation damage.
The primary trigger of the experiment was a coincidence between signals from the BigCal and from the HMS
 within a $\pm 50$ ns timing window.






In $\pi^{0}$ photoproduction, the meson decays into two photons directly following its production.
The minimum opening angle between these two decay photons 
corresponds to the two photons sharing the energy of the $\pi^{0}$ equally in the lab frame.
As the opening angle increases, one photon will take more energy from the $\pi^{0}$ and its track will be 
closer to the incident $\pi^{0}$ track direction. Either of the $\pi^{0}$ decay photons with energy greater
than the BigCal hardware energy threshold (set typically at about half the \textit{ep} elastic scattered electron 
energy) hitting the BigCal will produce a BigCal trigger. If the event was in coincidence with
 a proton in the HMS, it was recorded.
In two kinematic settings where the electron beam energy was 5.71 GeV, the BigCal coincidence acceptance with the HMS was large enough 
to detect both photons.
These data with lower statistics were also analyzed and the 
results were found to be consistent with the ``one photon detected'' results.
In this paper, only the ``one photon detected'' results will be shown.

To identify $\pi^{0}$ events when one photon was detected in the BigCal, the $\pi^{0}$ decay photon energy predicted from
 the proton angle, momentum and the $\pi^{0}$ decay photon angle was compared with the energy measured in the BigCal. 
A good linear correlation was seen between the measured and predicted energies.
We applied a 3$\sigma$ cut on the ratio of the measured and predicted photon energy to identify the  $\pi^{0}$ events. 
The major background events in the $\pi^{0}$ photoproduction channel come from the \textit{ep} elastic 
radiative tail and from random coincidence events. 
To reduce random background, a 3$\sigma$ cut around the BigCal and HMS coincidence time peak was applied. 
The \textit{ep} elastic radiation tail contamination was estimated by comparing the data to Monte Carlo simulation. 
Background events came from heavier meson photoproduction and multiple $\pi^{0}$ photoproduction were also estimated by the simulation. Only the data near the
Bremsstrahlung endpoint with less than 1.0\% contamination from these two types of reactions were kept in the analysis.

Figure \ref{pi0id} shows the distribution of the predicted $\pi^{0}$ decay photon energy $E_{calo}'$. 
 The left vertical dashed line indicates the hardware energy threshold of BigCal, 
and the right vertical dashed line 
indicates the $E_{calo}'$ upper limit selected to optimize the signal to background ratio and statistics. 
Events between the two vertical dashed lines were selected and used in the analysis.
At an incident photon energy of 3.951 GeV, the elastic background contamination ratio in the selected range of  $E_{calo}'$  is 1.2\%, and the random 
background contamination is 1.5\% after all cuts were applied. The polarization components of both 
kinds of background were studied separately and corrections were applied to the final results.

Elastic events were used to calibrate the FPP analyzing power and determine the instrumental asymmetry at each kinematic setting. 
With the knowledge of the beam polarization and of the spin precession in the HMS \cite{COSY}, 
polarization transfer in the \textit{ep} elastic reaction allows the determination of both the $CH_{2}$ 
analyzing power and the ratio of the proton electromagnetic form factors. 
To take into account the proton momentum difference between elastic events and $\pi^{0}$ events, 
the analyzing power of $\pi^{0}$ events was obtained by correcting the \textit{ep} elastic results
according to the analyzing power momentum dependence \cite{Ay_pCH2}.
 As the induced polarization in \textit{ep} elastic scattering is zero in the one photon 
exchange mechanism, 
the instrumental asymmetry could be extracted by Fourier analysis of the helicity sum spectrum
 of \textit{ep} elastic events. 
 The same cut on hit position of the protons in the focal plane of \textit{ep} elastic events was applied to 
the $\pi^{0}$ events to make sure the calibrated analyzing power and the instrumental asymmetry are valid. 
This cut also further suppressed the heavier meson (\textit{e.g.}, $\eta$) production contribution to the data by requiring higher proton momentum in the HMS. 
After all these calibrations were done, the induced 
and transferred polarization components of the proton in $\pi^{0}$ photoproduction at the target were 
extracted by the maximum likelihood method described in Ref. \cite{Puckett10}.

The high statistics of $\pi^{0}$ events allows us to divide 
the data into several incident photon energy bins.
The bin size was selected to be greater than the reconstructed incident photon energy resolution and
to keep enough events to calculate the polarization components in each bin. 
Systematic uncertainties were estimated by analyzing the sensitivity of
 the polarization components to background corrections, 
the beam polarization, the instrumental asymmetry, 
the analyzing power calibration and the tracking reconstruction systematics for each bin. 
For the polarization transfer components, the uncertainties from the \textit{ep} elastic 
background estimation are dominant because the polarization
 are very different in \textit{ep} elastic events.
 The systematic uncertainties of the induced polarization component are dominated by 
the instrumental asymmetry correction. 
Overall, the systematic uncertainties are less than $\pm$0.026 for the polarization transfer components
and do not exceed $\pm$0.034 for the induced polarization component.

The results are listed in Table \ref{result_table}.
No induced polarization data for the last kinematics in the table are available
 because the spin precession inside the HMS at this setting
leads to very large systematic uncertainties.
The lab coordinate system is defined by \^{z} = \^{k}$_{proton}$/$|$\^{k}$_{proton}$$|$, 
\^{y} = \^{k}$_{proton}$$\times$\^{k}$_{\gamma}$/$|$\^{k}$_{proton}$$\times$\^{k}$_{\gamma}$ $|$ and \^{x} = \^{y}$\times$\^{z}, 
where \^{k}$_{proton}$ (\^{k}$_{\gamma}$) is the recoil proton (incident photon) momentum.
$C_{z}^{lab}$, $P$ and $C_{x}^{lab}$ are the longitudinal, 
the induced (along \^{y}) and the transverse polarization components in the lab system, respectively.

 


Several theoretical models predict the polarization observables in the
$^{1}H(\vec{\gamma},\vec{p})\pi^{0}$ reaction;
 they are partial-wave analyses SAID \cite{PhysRevC.66.055213} and MAID \cite{MAID07} ($E_{\gamma} \le 1.65 \mbox{ GeV}$),
 a quark model sub-process calculation by Afanasev 
\textit{et al.} \cite{Afanasev1997393}, and a pQCD prediction from Farrar  \textit{et al.} 
\cite{Farrar1991655}.

In SAID, both an energy-dependent and a set of energy-independent
partial-wave analyses of single-pion 
photoproduction data were performed.
The latest SP09 \cite{SAID09} solution extends from threshold to 
2.7 GeV of incident photon energy in the laboratory. 


Assuming helicity conservation, the induced polarization P and
the transverse polarization transfer $C_{x}^{c.m.}$ in pion photoproduction
are zero. 
From pQCD scaling arguments, 
the longitudinal polarization transfer $C_{z}^{c.m.}$ is constant at fixed $\theta_{\pi}^{c.m.}$,
but HHC alone does not determine the value of this constant. 

\begin{figure*} [ht]
  \begin{center}
    \includegraphics[width=0.93\textwidth,height=.48\textwidth, angle=0,trim = 5mm 0mm 10mm 5mm, clip=]{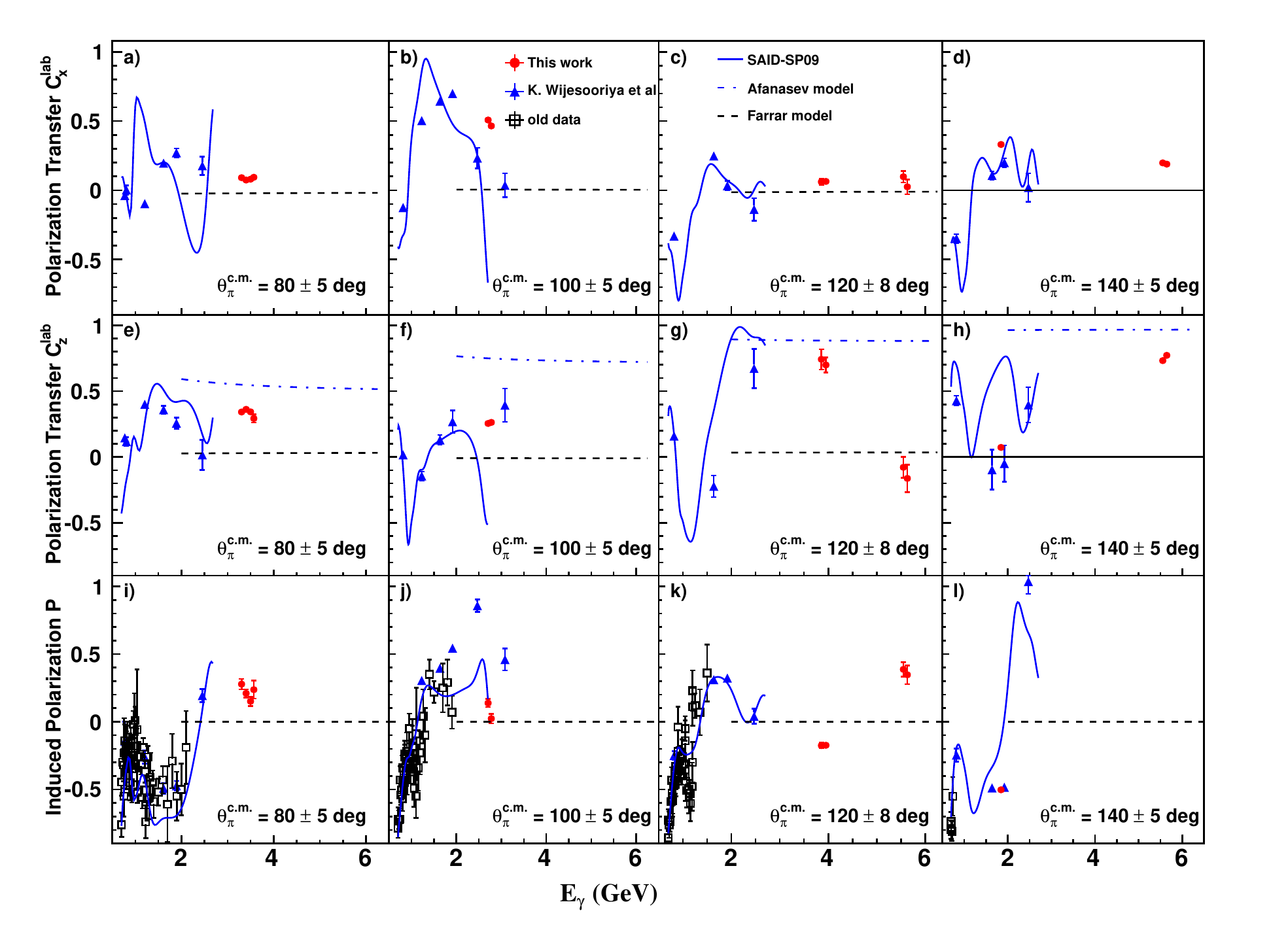}
  \end{center}
  \caption{Top to bottom: polarization transfer $C_{x}^{lab}$, $C_{z}^{lab}$, 
and induced polarization $P$ in the lab frame.
Left to right: different angles of $\pi^{0}$ in c.m. frame.
The ``old data'' could be found in the SAID data base \cite{PhysRevC.66.055213}. 
The three curves labeled Afanasev  model \cite{Afanasev1997393}, Farrar model \cite{Farrar1991655} and SAID SP09 \cite{SAID09} are described in the text.
Only the statistical uncertainties are shown.
\label{result_plot}}
\end{figure*}

Farrar \textit{et al.} predicted the helicity amplitudes for pion photoproduction by explicitly
calculating all lowest-order Feynman diagrams \cite{Farrar1991655}.
Several nucleon and pion wave functions were used in the calculation. 
The predicted cross sections are highly sensitive to the choices of wave functions
and they do not agree with the data in general. 
The calculated curves shown in Fig. \ref{result_plot} used asymptotic distribution amplitudes for both the
proton and the pion. 

Afanasev \textit{et al.} \cite{Afanasev1997393} used a pQCD approach to 
calculate the longitudinal polarization $C_{z}^{c.m.}$ of meson photoproduction in the limit
$x_{Bjorken}\rightarrow 1$.
This model assumes helicity conservation and that the pQCD approach is justified for 
high meson transverse momentum. 


Figure \ref{result_plot} presents the comparison of the new Hall C results with data from previous experiment and the available models.
Not all the data of \cite{gilman02} are shown in the figure.
 The theoretical predictions are calculated for the given $\pi^{0}$ c.m. angles shown in the panels
 and have been converted from the c.m. frame to the lab frame. 
In the lower incident photon energy regime ($E_{\gamma}<2.7$ GeV), 
these new data agree with the world data except for the induced 
polarization in Fig. \ref{result_plot} j.  
A strong $\theta_{\pi}^{c.m.}$ dependence for $P$ at $E_{\gamma} = 2.5$ GeV was found in the previous measurement \cite{gilman02}. 
The polarization dependence on $\theta_{\pi}^{c.m.}$ at $E_{\gamma} = 2.7$ GeV is studied, the results show
a compatible oscillation comparing to \cite{gilman02} for $P$ and $C_{x}^{lab}$.
This discrepancy very likely comes from the difference 
in $\theta_{\pi}^{c.m.}$ between the new data and the previous measurement. 
While the SAID model gives good overall predictions for energies lower than 3 GeV, it disagrees with the data in Fig. \ref{result_plot} panels a), j) and h); 
this can be understood since above 1 GeV the multipoles are still under-constrained in the model. 
For the larger incident photon energies ($E_{\gamma}>3.0$  GeV),
the new data are the first measurements at the given $\theta_{\pi}^{c.m.}$. 
The results still show strong energy dependence in $C_{z}^{lab}$ and P at 120 degrees, and a strong
angle dependence in $C_{z}^{lab}$ at $E_{\gamma} \approx 5.6$ GeV.
 Such behavior was not predicted by the models based on HHC.
It appears, based on our few examples, that the strong kinematic dependences in
the SAID fit at low energies continue up to 5.6 GeV. 


To conclude, the precise new polarization data for $\pi^{0}$ photoproduction from the proton presented here
extend the world data set to 
$E_{\gamma} = 5.6$ GeV. In the lower energy region, the new data are in good agreement with previous measurements and the SAID predictions. 
The new data for $E_{\gamma} < 2.7$ GeV together with data from MAMI-C \cite{MAMIB} will give further constraint on the multipole fit above 1 
GeV. 
At higher energy, the new data show no evidence of HHC up to $E_{\gamma} = 5.6$ GeV. 
Furthermore, the polarization transfer components vary drastically as a function of 
$\theta_{\pi}^{c.m.}$ at $E_{\gamma} \approx 5.6$ GeV, and this is
 not predicted by any theoretical model. 
The high energy data may allow interpretation in terms of the quark handbag mechanism, 
providing access to polarization-dependent Generalized Parton Distributions, 
as discussed in  \cite{Huang:2003jd}, \cite{Afanasev:1998ym}. 
More theoretical predictions would be highly 
desirable and the interpretation of the data would help achieve a complete understanding of the mechanism of this reaction.


We thank A. Afanasev for discussions of his model, 
and acknowledge the Hall C technical staff and the Jefferson Lab Accelerator Division
for their outstanding support during the experiment. 
This work was supported in part by the U.S. Department of Energy,
the U.S. National Science Foundation, the Italian Institute for Nuclear research, the French Commissariat \`{a}
l'Energie Atomique (CEA) and the Centre National de la Recherche Scientifique (CNRS), and the Natural Sciences and
Engineering Research Council of Canada. This work is supported by DOE contract DE-AC05-06OR23177, under
which Jefferson Science Associates, LLC, operates the Thomas Jefferson National Accelerator Facility.

\bibliography{main}

\newcommand{\noopsort}[1]{} \newcommand{\printfirst}[2]{#1}
  \newcommand{\singleletter}[1]{#1} \newcommand{\switchargs}[2]{#2#1}
\begin{thebibliography}{29}
\expandafter\ifx\csname natexlab\endcsname\relax\def\natexlab#1{#1}\fi
\expandafter\ifx\csname bibnamefont\endcsname\relax
  \def\bibnamefont#1{#1}\fi
\expandafter\ifx\csname bibfnamefont\endcsname\relax
  \def\bibfnamefont#1{#1}\fi
\expandafter\ifx\csname citenamefont\endcsname\relax
  \def\citenamefont#1{#1}\fi
\expandafter\ifx\csname url\endcsname\relax
  \def\url#1{\texttt{#1}}\fi
\expandafter\ifx\csname urlprefix\endcsname\relax\def\urlprefix{URL }\fi
\providecommand{\bibinfo}[2]{#2}
\providecommand{\eprint}[2][]{\url{#2}}

\bibitem[{\citenamefont{Brodsky and Farrar}(1973)}]{Brodsky1973}
\bibinfo{author}{\bibfnamefont{S.~J.} \bibnamefont{Brodsky}} \bibnamefont{and}
  \bibinfo{author}{\bibfnamefont{G.~R.} \bibnamefont{Farrar}},
  \bibinfo{journal}{Phys. Rev. Lett.} \textbf{\bibinfo{volume}{31}},
  \bibinfo{pages}{1153} (\bibinfo{year}{1973}).

\bibitem[{\citenamefont{Brodsky and Lepage}(1981)}]{Brodsky1981}
\bibinfo{author}{\bibfnamefont{S.~J.} \bibnamefont{Brodsky}} \bibnamefont{and}
  \bibinfo{author}{\bibfnamefont{G.~P.} \bibnamefont{Lepage}},
  \bibinfo{journal}{Phys. Rev. D} \textbf{\bibinfo{volume}{24}},
  \bibinfo{pages}{2848} (\bibinfo{year}{1981}).

\bibitem[{\citenamefont{Anderson et~al.}(1976)}]{Anderson1976}
\bibinfo{author}{\bibfnamefont{R.~L.} \bibnamefont{Anderson}}
  \bibnamefont{et~al.}, \bibinfo{journal}{Phys. Rev. D}
  \textbf{\bibinfo{volume}{14}}, \bibinfo{pages}{679} (\bibinfo{year}{1976}).

\bibitem[{\citenamefont{Bochna et~al.}(1998)}]{Bochna1998}
\bibinfo{author}{\bibfnamefont{C.}~\bibnamefont{Bochna}} \bibnamefont{et~al.},
  \bibinfo{journal}{Phys. Rev. Lett.} \textbf{\bibinfo{volume}{81}},
  \bibinfo{pages}{4576} (\bibinfo{year}{1998}).

\bibitem[{\citenamefont{Belz et~al.}(1995)}]{Belz1995}
\bibinfo{author}{\bibfnamefont{J.~E.} \bibnamefont{Belz}} \bibnamefont{et~al.},
  \bibinfo{journal}{Phys. Rev. Lett.} \textbf{\bibinfo{volume}{74}},
  \bibinfo{pages}{646} (\bibinfo{year}{1995}).

\bibitem[{\citenamefont{Freedman et~al.}(1993)}]{Freedman1993}
\bibinfo{author}{\bibfnamefont{S.~J.} \bibnamefont{Freedman}}
  \bibnamefont{et~al.}, \bibinfo{journal}{Phys. Rev. C}
  \textbf{\bibinfo{volume}{48}}, \bibinfo{pages}{1864} (\bibinfo{year}{1993}).

\bibitem[{\citenamefont{Schulte et~al.}(2001)}]{Schulte2001}
\bibinfo{author}{\bibfnamefont{E.~C.} \bibnamefont{Schulte}}
  \bibnamefont{et~al.}, \bibinfo{journal}{Phys. Rev. Lett.}
  \textbf{\bibinfo{volume}{87}}, \bibinfo{pages}{102302}
  (\bibinfo{year}{2001}).

\bibitem[{\citenamefont{Zhu et~al.}(2003)}]{PhysRevLett.91.022003}
\bibinfo{author}{\bibfnamefont{L.~Y.} \bibnamefont{Zhu}} \bibnamefont{et~al.}
  (\bibinfo{collaboration}{Jefferson Lab Hall A Collaboration}),
  \bibinfo{journal}{Phys. Rev. Lett.} \textbf{\bibinfo{volume}{91}},
  \bibinfo{pages}{022003} (\bibinfo{year}{2003}).

\bibitem[{\citenamefont{Chen et~al.}(2009)}]{PhysRevLett.103.012301}
\bibinfo{author}{\bibfnamefont{W.}~\bibnamefont{Chen}} \bibnamefont{et~al.}
  (\bibinfo{collaboration}{The CLAS Collaboration}), \bibinfo{journal}{Phys.
  Rev. Lett.} \textbf{\bibinfo{volume}{103}}, \bibinfo{pages}{012301}
  (\bibinfo{year}{2009}).

\bibitem[{\citenamefont{Klempt and Richard}(2010)}]{RevModPhys.82.1095}
\bibinfo{author}{\bibfnamefont{E.}~\bibnamefont{Klempt}} \bibnamefont{and}
  \bibinfo{author}{\bibfnamefont{J.-M.} \bibnamefont{Richard}},
  \bibinfo{journal}{Rev. Mod. Phys.} \textbf{\bibinfo{volume}{82}},
  \bibinfo{pages}{1095} (\bibinfo{year}{2010}).

\bibitem[{\citenamefont{Wijesooriya et~al.}(2002)}]{gilman02}
\bibinfo{author}{\bibfnamefont{K.}~\bibnamefont{Wijesooriya}}
  \bibnamefont{et~al.}, \bibinfo{journal}{Phys. Rev. C}
  \textbf{\bibinfo{volume}{66}}, \bibinfo{pages}{034614}
  (\bibinfo{year}{2002}).

\bibitem[{\citenamefont{Arndt et~al.}(2003)\citenamefont{Arndt, Strakovsky, and
  L.~Workman}}]{PhysRevC.67.048201}
\bibinfo{author}{\bibfnamefont{R.~A.} \bibnamefont{Arndt}},
  \bibinfo{author}{\bibfnamefont{I.~I.} \bibnamefont{Strakovsky}},
  \bibnamefont{and}
  \bibinfo{author}{\bibfnamefont{R.}~\bibnamefont{L.~Workman}},
  \bibinfo{journal}{Phys. Rev. C} \textbf{\bibinfo{volume}{67}},
  \bibinfo{pages}{048201} (\bibinfo{year}{2003}).

\bibitem[{\citenamefont{Huang et~al.}(2004)\citenamefont{Huang, Jakob, Kroll,
  and Passek-Kumericki}}]{Huang:2003jd}
\bibinfo{author}{\bibfnamefont{H.~W.} \bibnamefont{Huang}},
  \bibinfo{author}{\bibfnamefont{R.}~\bibnamefont{Jakob}},
  \bibinfo{author}{\bibfnamefont{P.}~\bibnamefont{Kroll}}, \bibnamefont{and}
  \bibinfo{author}{\bibfnamefont{K.}~\bibnamefont{Passek-Kumericki}},
  \bibinfo{journal}{Eur. Phys. J.} \textbf{\bibinfo{volume}{C33}},
  \bibinfo{pages}{91} (\bibinfo{year}{2004}).

\bibitem[{\citenamefont{Ji et~al.}(2003)\citenamefont{Ji, Ma, and
  Yuan}}]{PhysRevLett.90.241601}
\bibinfo{author}{\bibfnamefont{X.}~\bibnamefont{Ji}},
  \bibinfo{author}{\bibfnamefont{J.-P.} \bibnamefont{Ma}}, \bibnamefont{and}
  \bibinfo{author}{\bibfnamefont{F.}~\bibnamefont{Yuan}},
  \bibinfo{journal}{Phys. Rev. Lett.} \textbf{\bibinfo{volume}{90}},
  \bibinfo{pages}{241601} (\bibinfo{year}{2003}).

\bibitem[{\citenamefont{Dutta and Gao}(2005)}]{PhysRevC.71.032201}
\bibinfo{author}{\bibfnamefont{D.}~\bibnamefont{Dutta}} \bibnamefont{and}
  \bibinfo{author}{\bibfnamefont{H.}~\bibnamefont{Gao}},
  \bibinfo{journal}{Phys. Rev. C} \textbf{\bibinfo{volume}{71}},
  \bibinfo{pages}{032201} (\bibinfo{year}{2005}).

\bibitem[{\citenamefont{Puckett et~al.}(2010)}]{Puckett10}
\bibinfo{author}{\bibfnamefont{A.~J.~R.} \bibnamefont{Puckett}}
  \bibnamefont{et~al.}, \bibinfo{journal}{Phys. Rev. Lett.}
  \textbf{\bibinfo{volume}{104}}, \bibinfo{pages}{242301}
  (\bibinfo{year}{2010}).

\bibitem[{\citenamefont{Meziane et~al.}(2011)}]{Meziane11}
\bibinfo{author}{\bibfnamefont{M.}~\bibnamefont{Meziane}} \bibnamefont{et~al.},
  \bibinfo{journal}{Phys. Rev. Lett.} \textbf{\bibinfo{volume}{106}},
  \bibinfo{pages}{132501} (\bibinfo{year}{2011}).

\bibitem[{\citenamefont{Hauger et~al.}(2001)}]{MollerNIM}
\bibinfo{author}{\bibfnamefont{M.}~\bibnamefont{Hauger}} \bibnamefont{et~al.},
  \bibinfo{journal}{Nucl. Instrum. Methods {A}} \textbf{\bibinfo{volume}{462}},
  \bibinfo{pages}{382} (\bibinfo{year}{2001}).

\bibitem[{\citenamefont{Olsen and Maximon}(1959)}]{Olsen1958}
\bibinfo{author}{\bibfnamefont{H.}~\bibnamefont{Olsen}} \bibnamefont{and}
  \bibinfo{author}{\bibfnamefont{L.~C.} \bibnamefont{Maximon}},
  \bibinfo{journal}{Phys. Rev.} \textbf{\bibinfo{volume}{114}},
  \bibinfo{pages}{887} (\bibinfo{year}{1959}).

\bibitem[{\citenamefont{Blok et~al.}(2008)}]{Blok2008}
\bibinfo{author}{\bibfnamefont{H.}~\bibnamefont{Blok}} \bibnamefont{et~al.},
  \bibinfo{journal}{Phys. Rev. C} \textbf{\bibinfo{volume}{78}},
  \bibinfo{pages}{045202} (\bibinfo{year}{2008}).

\bibitem[{\citenamefont{Makino and Berz}(1999)}]{COSY}
\bibinfo{author}{\bibfnamefont{K.}~\bibnamefont{Makino}} \bibnamefont{and}
  \bibinfo{author}{\bibfnamefont{M.}~\bibnamefont{Berz}},
  \bibinfo{journal}{Nucl. Instrum. Methods A} \textbf{\bibinfo{volume}{427}},
  \bibinfo{pages}{338} (\bibinfo{year}{1999}).

\bibitem[{\citenamefont{Azghirey et~al.}(2005)}]{Ay_pCH2}
\bibinfo{author}{\bibfnamefont{L.~S.} \bibnamefont{Azghirey}}
  \bibnamefont{et~al.}, \bibinfo{journal}{Nucl. Instrum. Methods A}
  \textbf{\bibinfo{volume}{538}}, \bibinfo{pages}{431} (\bibinfo{year}{2005}).

\bibitem[{\citenamefont{Arndt et~al.}(2002)\citenamefont{Arndt, Briscoe,
  Strakovsky, and Workman}}]{PhysRevC.66.055213}
\bibinfo{author}{\bibfnamefont{R.~A.} \bibnamefont{Arndt}},
  \bibinfo{author}{\bibfnamefont{W.~J.} \bibnamefont{Briscoe}},
  \bibinfo{author}{\bibfnamefont{I.~I.} \bibnamefont{Strakovsky}},
  \bibnamefont{and} \bibinfo{author}{\bibfnamefont{R.~L.}
  \bibnamefont{Workman}}, \bibinfo{journal}{Phys. Rev. C}
  \textbf{\bibinfo{volume}{66}}, \bibinfo{pages}{055213}
  (\bibinfo{year}{2002}).

\bibitem[{\citenamefont{Drechsel et~al.}(2007)\citenamefont{Drechsel, Kamalov,
  and Tiator}}]{MAID07}
\bibinfo{author}{\bibfnamefont{D.}~\bibnamefont{Drechsel}},
  \bibinfo{author}{\bibfnamefont{S.}~\bibnamefont{Kamalov}}, \bibnamefont{and}
  \bibinfo{author}{\bibfnamefont{L.}~\bibnamefont{Tiator}},
  \bibinfo{journal}{The European Physical Journal A - Hadrons and Nuclei}
  \textbf{\bibinfo{volume}{34}}, \bibinfo{pages}{69} (\bibinfo{year}{2007}),
  ISSN \bibinfo{issn}{1434-6001}.

\bibitem[{\citenamefont{Afanasev et~al.}(1997)\citenamefont{Afanasev, Carlson,
  and Wahlquist}}]{Afanasev1997393}
\bibinfo{author}{\bibfnamefont{A.}~\bibnamefont{Afanasev}},
  \bibinfo{author}{\bibfnamefont{C.~E.} \bibnamefont{Carlson}},
  \bibnamefont{and}
  \bibinfo{author}{\bibfnamefont{C.}~\bibnamefont{Wahlquist}},
  \bibinfo{journal}{Physics Letters B} \textbf{\bibinfo{volume}{398}},
  \bibinfo{pages}{393 } (\bibinfo{year}{1997}).

\bibitem[{\citenamefont{Farrar et~al.}(1991)\citenamefont{Farrar, Huleihel, and
  Zhang}}]{Farrar1991655}
\bibinfo{author}{\bibfnamefont{G.~R.} \bibnamefont{Farrar}},
  \bibinfo{author}{\bibfnamefont{K.}~\bibnamefont{Huleihel}}, \bibnamefont{and}
  \bibinfo{author}{\bibfnamefont{H.}~\bibnamefont{Zhang}},
  \bibinfo{journal}{Nuclear Physics B} \textbf{\bibinfo{volume}{349}},
  \bibinfo{pages}{655 } (\bibinfo{year}{1991}), ISSN \bibinfo{issn}{0550-3213}.

\bibitem[{\citenamefont{Dugger et~al.}(2009)}]{SAID09}
\bibinfo{author}{\bibfnamefont{M.}~\bibnamefont{Dugger}} \bibnamefont{et~al.}
  (\bibinfo{collaboration}{CLAS Collaboration}), \bibinfo{journal}{Phys. Rev.
  C} \textbf{\bibinfo{volume}{79}}, \bibinfo{pages}{065206}
  (\bibinfo{year}{2009}).

\bibitem[{\citenamefont{Sikora}(2011)}]{MAMIB}
\bibinfo{author}{\bibfnamefont{M.}~\bibnamefont{Sikora}},
  \bibinfo{type}{{Ph.D.} thesis}, \bibinfo{school}{Edinburgh Univ.}
  (\bibinfo{year}{2011}).

\bibitem[{\citenamefont{Afanasev}(1998)}]{Afanasev:1998ym}
\bibinfo{author}{\bibfnamefont{A.~V.} \bibnamefont{Afanasev}},
  \bibinfo{journal}{arXiv:hep-ph/9808291}  (\bibinfo{year}{1998}).

\end{thebibliography}

\end{document}